\documentstyle[12pt,aasms4]{article}

\begin{document}

\title{Host galaxies of z$\sim$4.7 QSOs
\footnote{Based on observations obtained at the Gemini Observatory, which is
operated by the Association of Universities for Research in Astronomy, Inc.,
under a cooperative agreement with the NSF on behalf of the Gemini partnership:
the National Science Foundation (United States), the Particle Physics and
Astronomy Research Council (United Kingdom), the National Research Council
(Canada), CONICYT (Chile), the Australian Research Council
(Australia), CNPq (Brazil) and CONICET (Argentina)} } 

\author{J.B. Hutchings}
\affil{Herzberg Institute of Astrophysics, NRC of Canada,\\ Victoria, B.C.
V8X 4M6, Canada; john.hutchings@nrc.ca}

\begin{abstract}

     2$\mu$ broad- and narrow-band imaging with the Gemini-N 
telescope of five z$\sim$4.7 QSOs, has resolved both the host galaxies 
and [O II] emission-line gas. The resolved fluxes of the host galaxies
fall within the extrapolated spread of the K-z relationship for radio
galaxies at lower redshifts, and their resolved morphology is irregular. 
The [O II] images indicate knots coincident with many continuum features
and also some bright jet-like features near the nucleus. The line emission
total fluxes indicate overall equivalent widths of 5 to 10\AA~ at rest
wavelengths. Two of the QSOs are in a local environment of faint galaxies 
of similar magnitude to the hosts, and three have nearby galaxies with 
excess narrow-band flux, which would be [O~II] if they are at the QSO redshift. 

\end{abstract} 


\section{Introduction and observations}
  
     The host galaxies of high redshift QSOs are of interest as they
are considered to be massive galaxies with high abundances and massive
central black holes in the early universe. It has been found that at
redshift 2 and higher, QSO host galaxies are large, luminous, and the sites
of star-formation and galaxy merging. There have been several investigations 
of QSO host galaxies in the redshift range 2 to 2.5 (e.g. Ridgway et al
2001, Kukula et al 2001, Hutchings et al 2002). There have been studies of
radio galaxies, which are easier to resolve, to redshifts 4 and higher
(e.g.  van Breugel et al 1998, Inskip et al 2001, Jarvis et al 2001,
De Breuck et al 2002). Compared with radio galaxies, QSO hosts are more
difficult to study because of the bright nuclear light source, but recent
surveys, in particular the SDSS, have greatly increased the number of QSOs
as known high redshift objects. This paper describes observations
intended to detect and characterize some QSO hosts at redshifts near 4.7.

    Most of the work on higher redshift QSOs has been done with HST or
ground-based AO systems, to reduce the light scattered by the QSO nucleus.
While these studies have found that there are compact spheroidal
stellar populations around the nuclei (see above references), there are 
also indications that the host galaxies have faint light extended over 
several arcsec (e.g. Hutchings 1995). Assuming that the hosts are still 
large and luminous at redshifts 4.7,
the main requirement was of large light-gathering power, since redshift
dimming will be large. Subarcsecond image quality is still very desirable,
to avoid too much contamination by the bright unresolved nucleus. Finally,
in order to learn about both stars and gas in the galaxies, the observations
were made in broad-band and narrow-band at the redshifted wavelength of 
the [O II] emission line, characteristic of star-formation. Thus, the
rest wavelengths are near 3700\AA, and the observations made near 2 microns.
Similar narrow-band imaging of the redshifted [O III] lines is described by
Hutchings, Morris, and Crampton (2001).

   The QSOs were selected from a high redshift list of Djorgovski 
(http://www.astro.caltech.edu/$\sim$george/z4.qsos) based mainly
on the SDSS (e.g. Anderson et al 2001, Zheng et al 2000). The sample
selected has redshifts that place the [O~II] line inside the bandpass
of a NIRI narrow-band filter. Objects were selected from this short list
by the Gemini service observers, to fit their schedules and weather conditions.
Thus, the sample is essentially random in all properties other than redshift
being close to 4.7, and the magnitude limits of the SDSS.

     The observations were carried out on the Gemini N telescope
during the first half of
2002, with total exposures of 40 minutes in both broad- and
narrow-band filters. The details are given in Table 1, along
with properties of the QSOs, including the R/r magnitude and QSO names
from Djorgovski's list, and some measured quantities. As the
observations were carried out by service observing over a number 
of nights, the conditions varied somewhat between them. The FWHM
and limiting surface brightness are shown for all objects, and
lie in reasonably compact ranges of values. We refer below to the
narrow- and broad-band observations as NB and BB respectively.

     The analysed images were derived from dithered exposures, 
processed using standard techniques. Details of this process for 
other ground-based observations are given in 
Hutchings et al (1999), and Steinbring, Crampton, and Hutchings (2002).
The field size is approximately 
2 arcmin, which included suitable PSF stars in all cases, and enabled 
some analysis of the faint galaxy counts in the QSO neighbourhood.

\section{Data and measurements}

   The QSOs were identified by their position and catalogued magnitude
using finding charts from the Digitised Sky Survey. With one 
exception noted below, this process was simple and unambiguous.

   For each QSO, we derived a PSF from a nearby bright and single star.
In general there was only one suitable PSF star (one field had only a
star with a fainter companion that had to be edited out), so that combining
stars to improve the PSF was not possible. However, a few other fainter stars
were plotted to be sure that the principal one is unresolved and unblended.

Azimuthally averaged luminosity plots were made (using IRAF task \it ellipse
\rm) for the PSF star, the QSO, and in some cases, the nearest resolved faint 
galaxy which might be a companion. These plots are shown in Figure 1, 
and demonstrate clearly where QSO images are resolved. Similar plots were 
made for a few other stars or compact galaxies in the images that have
similar signal and lack of nearby companions, to the QSOs. These plots 
lie close to the PSF, and thus suggest that the resolved light around the
QSOs is associated (i.e. `host' galaxy), rather than due to faint 
background objects that do not show up in the PSFs, which are derived 
from bright stars.

The resolved flux was calculated from the plots in Figure 1, and also 
with a range of PSF scaling factors 
that leave a difference profile that extends approximately linearly to zero 
radius. This does not include the possibility of an unresolved compact
core of the host galaxy.

   In addition, the PSF star image was scaled, shifted, and subtracted 
from the QSO image, so as
to leave the weakest residual image that has a smoothly rising flux to 
the centre. These PSF-subtracted images were also measured as the resolved
host galaxies.

Finally, the broad-band image was scaled and subtracted from the narrow-band,
to isolate resolved line emission. In some cases the image FWHM differs 
between the broad and narrow band images, so that the residuals are affected
by this. We discuss the QSOs individually below, and the errors are
discussed in the final section. In these measurements and plots, we removed
objects that appear to be non-associated galaxies in the line of sight, so 
that they do not lead to overestimation of the host galaxy flux. In no
case do such objects amount to an appreciable fraction of what was 
measured and shown in Table 1.

   Other galaxy fluxes in the fields were measured in both broad and 
narrow- band images. These were measured with the IRAF task \it imexam \rm
set with a radius of 1.2 arcsec. Thus, larger galaxies and images with
larger FWHM give incomplete flux values, but the intention was to compare 
the ratio of narrow to broad band flux for the compact objects, consistently
across the fields for each object. Objects with line emission in the
narrow-band image will have a higher ratio of narrow to broad band flux.
In all cases we find this is true of the QSO, and there are other
faint emission line candiates that may be companions, as we note
in the sections on individual QSOs.

   The calibration data were used to obtain approximate magnitudes for all
measurements, where the uncertainty lies in some observing conditions 
being somewhat non-photometric. The limiting magnitude is fairly consistent
among the images, and we have measured broad-band galaxy images down to 
about K=22. We have counted faint galaxies to this limit in all fields,
centred on the QSO, and in one corner of the image as a comparison
subfield. These are shown in Table 2. Galaxies brighter than the QSO are
not counted as they must be foreground objects. However, their numbers are
shown as general information about the field.

\subsection{SDSS1321+0038}

  The positional ID is slightly doubtful. The QSO is assumed to be the
relatively unresolved object of the expected magnitude, although a nearby
brighter resolved galaxy is at the nominal position. There is a bright PSF 
star very nearby too. The field is uncrowded and there is no obvious population
or grouping of faint galaxies. The flux measures for all three objects 
indicate no significant additional NB emission from the QSO, within the 
10\% scatter of the measures. The companion galaxy is 
actually fainter by about 8\% in the NB image.

   The luminosity profiles show marginally resolved flux in the BB images
but indicate clearly resolved flux in the NB outside of 1.5 arcsec radius.
The brighter companion galaxy is well resolved at all radii.
The companion lies only 5 arcsec from the QSO, and its light effects
are measurable as close as 3.5 arcsec. Figure 2 shows both images.
The companion slightly affects measures of the NB image, which shows a 
curved filament in this direction, and the resolved NB flux is lower 
than indicated by the ellipse-fit profile in Figure 1. Allowing for this, 
its BB magnitude is 20.7.

   The immediate environment ($\pm$15") of the QSO is low in faint galaxy
counts, compared with the whole field. Twelve galaxies are 2.5 - 3.5 
magnitudes fainter than the QSO (40\%), and three of these have the lowest
NB/BB ratios (2.0 - 2.3), compared with 3.5 for the QSO and a mean of 4.2 
for the
other faint galaxies. Thus, they are candidates for companions to the
QSO, with strong [O II] emission.

\subsection{PC1415+3408}

    The NB star images are quite elliptical and the BB slightly
elliptical. For both, the mean FWHM are the worst of the dataset. There is only
one good PSF star, and it has a nearby companion that has to be edited out.
The result of all this is that the QSO is unresolved out to about 2
arcsec radius. Beyond this, there is some faint resolved structure in both
NB and BB. There is a faint companion at this distance and is likely what 
we see. The NB image has some jet-like extension/blob closer in (0.9 arcsec)
along the short axis of the guiding ellipse, which does not show up in the
luminosity plot. The NB image has 2 bright pixels near the nucleus that 
do not fit the profile. The BB image has similar effect but may be due 
to a mismatch of the edited PSF in the subtraction.
The NB image has 0.1mag more flux than the continuum image, assuming all
images are photometric.

   The galaxy count close to the QSO is slightly lower than the frame
average, and the galaxy count is the lowest of the 5 fields. There are 11
galaxies (52\%) more than ten times fainter than the QSO,
and two have unusually high NB/BB ratio (1.6 and 2.2 compared with 3.2
for the QSO and a mean of 4.5 for the other faint galaxies).

\subsection{GB1428+4217}

  The QSO has 15\% extra signal in the NB image, and the magnitude is 0.1 
brighter. It lies in a field of crowded small faint galaxies (about 60 in the
image to K=19.5, and many times that down to K=21.0). 

   The radial plots were done after editing out the nearest companion, about
3.1 arcsec away. The BB image is marginally resolved beyond 1.8 arcsec,
and clearly resolved beyond 2.4 arcsec. The NB image shows resolved 
light beyond 1.5 arcsec.

   The broad-band image shows a very unsmooth structure of the light
resolved around the QSO. While the innermost arcsec is unresolved,
there is no symmetrical connection to the outer structures, in the total 
image or after PSF-subtraction. The QSO thus appears to be located in a central
unresolved blob surrounded by several radial wisps whose total flux is
given in Table 1. A few other wisps lie within 8 arcsec, unconnected to the
QSO, and which are not measured. A compact resolved brighter blob lies
3.5 arcsec away, unconnected to the wisps.

   The narrow-band images show an extension that is directed some 45$^o$
away from the line to the nearby companion. It extends to 2.4". There 
is also extended light close to the nucleus orthogonal to this. The 
PSF-subtracted NB image suggests some structure along a different axis
within the innermost arcsec, curving on one side into the outer extension,
and a linear inner feature that corresponds to the inner extended light.
The N-B subtracted image shows similar morphology in all these respects.
The good image quality, and similar FWHM in BB and NB images allow us to
have confidence in these morphological details. Figure 3 shows some
images of the QSO.

   The extended resolved NB light corresponds to mag 21.3, although the
NB/BB ratio suggests an equal additional amount of NB luminosity from 
the unresolved central region.

   There are many faint galaxies in the field, with two empty patches
away from the QSO. The galaxy density near the QSO is typical of the field,
and over this field has nearly twice the number of galaxies as the other QSOs.
The QSO is the brightest object, and most of the galaxies are faint enough
to be potential companions. Nine galaxies have N/B flux ratios that are
low enough to suggest [O II] emission at the QSO redshift (1.6 - 2.2,
compared with 3.2 for the QSO, and a mean of 4.5 for the other faint galaxies.
Four of these candidates are in the 9 galaxies closest to the QSO, and 3 
of those 9 are too faint to measure
NB flux. This seems a good candidate for a cluster associated with the QSO. 
 
\subsection{SDSS1451-0104}

  This QSO is discussed by Zheng et al (2000), along with a spectrum.
There is a broad absorption trough shortward of C IV, but separate from the
emission, so it may be a BAL. The Oxygen lines in the rest-frame UV are
strong, so we may expect [O II] to be strong.

   This object has a significant mismatch in resolution of the NB and BB
images, but the NB images are very good. The NB image has 16\% more signal
than the scaled BB image, using the average of a star and nearby faint
galaxy for comparison. There are few faint galaxies in the image, so no
grouping around the QSO is indicated. 

   The radial plots show the BB images is marginally resolved from a radius
of 0.9 arcsec, but with flux that falls almost as rapidly as the PSF.
Beyond 2.1", the BB image scale length is larger.
The NB profile shows clear resolved flux beyond 1.3 arcsec, with a relatively
large scale length. The images show some compact extended flux in the BB
image that looks like a broad tail on one side. The NB image shows some
extension in this direction, more extended, but also a brighter linear
feature at 120$^o$ to this, with some sharp curvature near the end. Figure 4
shows some images of the QSO.

    The QSO lies in a region less crowded with galaxies than the frame
average. There are 18 galaxies more than 10 times fainter than the QSO,
almost all of which lie in a group away from the QSO. None are bright
enough to measure the NB/BB flux ratio.

\subsection{SDSS1532-0039}

   This was observed several times, with different image quality. The QSO
lies centrally in a grouping of faint galaxies within the whole field imaged.
The worst
BB image was not used, and the others used both separately and combined. 
The NB images were better resolved. Both sets of NB and BB images gave 
different signal levels, so it was assumed that the lower signal ones 
non-photometric. The QSO has a nearby star and companion galaxy, both of
comparable brightness that were used for the PSF and photometric comparison. 
The NB QSO images were 23\% brighter in the NB than
the scaling for the star and companion, indicating this level of line emission
in the NB bandpass. This has an uncertainty of about 3\% from the star to
companion ratio spread.

   In the radial luminosity plots, the BB image shows no resolved
structure within 2.3 arcsec. Beyond that, faint flux is resolved to the 
limit of the image.
There is a companion galaxy some 3.5 arcsec away that was edited out of the
images for radial plots. This editing may affect the luminosity plot to a
small extent in the 2-4 arcsec radius range. Plots were done for the
companion by editing out the QSO, so similar considerations apply. The
companion is clearly resolved in all images down to radii of 0.35 arcsec.

The QSO NB images, on the other hand, show resolved structure clearly 
beyond 0.8 arcsec to the limiting signal at radius at 4.1 arcsec. 
Both NB images show extended structure as a jet-like feature, 
out to radius about 4 arcsec. The jet seems detached from the main image
and not quite radial, and details differ between the two NB images, 
which indicates the level of noise. Figure 2 shows broad- and narrow-band
images. The magnitude of the resolved NB structure,
in BB equivalent is 20.5, or some 5\% of the QSO flux. 

The BB images show linear radial structure similar in size and direction to 
the NB, but fainter and hard to measure. They also show a faint
companion centred at 3.5 arcsec in this direction (just outside the region
shown in Figure 2), with magnitude 22.5.

  The QSO lies in a group of galaxies with sky density above the frame average.
Apart from the 6 very bright galaxies, all galaxies are faint enough
to be possible QSO companions.  None of them has an unusual NB/BB flux
ratio, however.

\section{Discussion}

  While the broad-band images all have luminosity profiles that are
broader than the PSF at `large' radii (2 arcsec or more), the images do not
indicate a smooth centrally concentrated host galaxy. It appears more that
the QSO is located in one centrally located unresolved body within a
network of knots and filaments that extend to several arcsec. This distribution
is not found in randomly located faint objects in the field, and in any
case is derived from images cleaned of knots that may be unassociated
neighbours. Average and median images from all the QSOs have profiles that
show the same resolved light, again suggesting that this is not random faint
background objects. Thus, we may be seeing host galaxies that are in
early stages of assembly. 

   We have several ways to estimate the host galaxy flux. We can integrate the
resolved light under the PSF from the profile plots. We can extrapolate this
light to zero radius (but have no particular model to assume). Finally, we can
measure the flux remaining after our `best' PSF-subtraction (as described
above). The range of these estimates is less than one magnitude in all cases.
We have adopted a `best estimate' shown in the table, based on all
considerations, taking into account the image quality for each object. 
The differences between the resolved flux and the host estimates give a
good idea of the error bars to apply to the latter. The host values from
Table 1 are the values plotted in Figure 5. We note that the values fall 
among the high redshift galaxies from de Breuck et al (2002), or form a
reasonable extrapolation of them.
Thus, perhaps not surprisingly, we find that luminous QSOs at redshift 4.7
are found inside the most luminous galaxies, and that the stellar populations
will evolve along the K-z relationship.

  The [O II] line emission is resolved all five QSOs. In the luminosity
plots, this occurs at smaller radii than in the continuum images, and shows
up as small bright (some jet-like) structures in the NB-BB or even the
raw NB images. The overall NB/BB flux ratios indicate unresolved [O II] flux
as well, in all cases. The morphology of the [O II] corresponds to
regions of continuum flux, but the line emission occurs in smaller bright 
spots, as one would expect for regions with active star-formation. The
similar distribution of [O II] and continuum light further confirms the
picture of host galaxies in assembly. The total [O II] flux amounts to
a rest-frame equivalent width of 5 to 10\AA, which is typical of
star-forming galaxies, and also typical of QSO nuclear (or NLR) spectra.
However, we can see that the line emission is
localised to small regions within the host galaxy, where star-formation 
must be very active.  

   The fact that the QSOs are bright presumably indicates that massive 
black holes have evolved at this redshift. Whether the black holes are 
related to the central compact stellar population rather than the larger
assembling galaxy is an interesting issue in understanding the tight 
BH-bulge relation seen at lower redshifts. However, as these are very 
bright QSOs we must also question whether they are selected as those 
beamed at us, so that their total luminosity is lower than it appears to us.

  Finally, two of the QSOs live in more than usually crowded regions of 
faint galaxies, and in three cases there are a few faint companions that
may be in active star-formation if their NB/BB flux ratios do indicate 
[O II] emission at the QSO redshift. The flux ratios are consistent with this.
In this connection we note the group of emission-line objects found by
Venemans et al (2002) around a z=4.1 radio galaxy, and the high star-formation
rates discussed for z$\sim$3 Lyman break galaxies by Shapley et al (2001).

   I thank Wil van Breugel and Carlos De Breuck for providing their list 
of values for the K-z plot.

\clearpage
\begin{deluxetable}{lrlllrll}
\tablenum{1}
\tablecaption{NIR imaging of z$\sim$4.7 QSOs}
\tablehead{
\colhead{Name} &
\colhead{FWHM} &
\colhead{z} &
\colhead{Magnitudes} &
\colhead{Filters\tablenotemark{1}} &
\colhead{NB\%} &
\colhead{K$_{res}$\tablenotemark{2}} &
\colhead{K$_{host}$\tablenotemark{3}} \\
&\colhead{(")} && \colhead{R, K, K$_{lim}$\tablenotemark{4}} 
&\colhead{BB, NB} &&[O II]}
\startdata
1321+0038 &BB 0.77 &4.67 &21.5, 18.3, 25.3 &K', H2 1-0 &$<$10~ &20.7 &20.5\cr
          &NB 0.73 &&&&&20.7\cr
\cr
1415+3408 &BB 0.89 &4.60 &21.4, 18.4, 24.4 &Kshort, Kcont &10~ &20.1 &19.8\cr
          &NB   0.94 &&&&&--\cr
\cr
1428+4217 &BB 0.72 &4.715 &20.9, 17.4, 24.5 &K', H2 1-0  &15~ &20.5 &20.2\cr
          &NB 0.68 &&&&&21.3\cr
\cr
1451$-$0104 &BB 0.56 &4.672 &22.7, 18.4, 25.4 &K', H2 1-0  &16~ &20.0 &19.8\cr
          &NB 0.89 &&&&&20.7\cr   
\cr
1532$-$0039 &BB 1.10 &4.62 &21.2, 17.3, 23.5  &K', Kcont &23~ &19.7 &19.3\cr
           &0.73 \cr
           &0.83\cr
           &NB 0.68 &&&&&20.5\cr
           &0.53\cr
\enddata
\tablenotetext{1}{Broad- and Narrow-band filters used: see NIRI website}
\tablenotetext{2}{Of resolved flux beyond PSF in profiles}
\tablenotetext{3}{Estimated from PSF-removal and extrapolated flux profile}
\tablenotetext{4}{Limiting surface brightness of plots}
\end{deluxetable}

\begin{deluxetable}{lllll}
\tablenum{2}
\tablecaption{Faint galaxy counts}
\tablehead{
\colhead{Name} &
\colhead{60'} &\colhead{30'} &\colhead{Corner 30'} }
\startdata
1321+0038    &  28 (3)  &        3  &       8\cr
1415+3408     &21 (6)    &      3    &     5\cr
1428+4217      &45 (10)   &     12    &    12\cr
1451$-$0104     & 27 (5)     &     5     &    7\cr
1532$-$0039      &24 (6)     &     8      &   3\cr
\enddata
     Numbers in parentheses are galaxies brighter than the QSO

\end{deluxetable}

\clearpage

\centerline{References}

Anderson S.F., et al, AJ, 2001 (astro-ph/0103228)

De Breuck C., van Breugel W., Stanford S.A., Rottgering H., Miley G.,
Stern D., 2002, AJ, 123, 637

Hutchings J.B., 1995, AJ, 110, 994

Hutchings J.B., Morris S.L., Crampton D., 2001, AJ, 121, 80

Hutchings J.B., Crampton D., Morris S.L., Durand D., Steinbring E., 1999, 
AJ, 117, 1109

Hutchings J.B., Frenette D., Hanisch R.J., Mo J., Dumont P.J., Redding D.C.,
Neff S.G., 2002, AJ, 123, 2936

Inskip K.J., Best P.N., Longair M.S., MacKay D.J.C., 2002, MNRAS, 329, 277 
(astro-ph/0110054)

Kukula M.J., Dunlop J.S., McLure R.J., Miller L., Percival W.J., Baum S.A.,
O'Dea C.P., 2001, MNRAS, 326, 1533 (astro-ph/0010007)

Jarvis M.J. et al, 2001, MNRAS, 326, 1585

Ridgway S.E., Heckman T.M., Calzetti D., Lehnert M., 2001, ApJ, 550, 122 (astro-ph/0011330)

Shapley A.E., Steidel C.C., Adelberger K.L., Dickinson M., Giavalisco M.,
Pettini M., 2001, ApJ, 562, 95

Steinbring E., Crampton D., Hutchings J.B., 2002, ApJ, 569, 611

van Breugel W., Stanford S.A., Spinrad H., Graham J.H., 1998, ApJ, 502, 614

Venemans B.P., et al, 2002, ApJ, 569, L11

Zheng W., et al, AJ, 2000 (astro-ph/0005247)

\clearpage

\centerline{Captions to figures}

1. Luminosity plots of semi-major axes of best-fitting ellipses, for
QSO and PSF images. Each QSO has a pair of plots, for narrow-band and 
broad-band images. In some cases the profiles of nearby compact companions
are shown as dotted lines, shifted to compare their outer isophote plots.
The unshifted central values of these are shown as open circles. The
lower right panel shows the profile derived from the mean of all 5 QSO images,
scaled to the same signal levels, and the mean of their PSFs. Table 1
gives the surface brightnesses of the faintest QSO values plotted in each
panel.

2. Grey-scale images, 7 arcsec on a side. Top images are 1532-0039 and lower
images are 1321+0038; left side are broad-band and right side are narrow-band.  
The 1532-0039 images also show the nearby resolved companion, which is the
right-hand object. 1321+0038 has a bright star just off the frame to the left. 

3. Grey-scale images of 1428+4217. Top left: NB-PSF image; top right: 
NB-BB difference image; lower left NB image; lower right: BB image.
Each panel is 14" on a side. The images have been smoothed with a 0.14"
gaussian. Note the extended line emission in the upper
panels, and the positional correspondence between NB and BB faint outer
knots in the lower panels.
 
4. Grey-scale images of 1451-0104. Left panels are broad-band and right
panels are narrow-band images. The upper panels are 7" on a side and the
lower are 14". The upper panels show the inner QSO structure and
the lower panels show faint outer structures and knots. The images have been
smoothed with a 0.14" gaussian.

5. K-z plot for radio galaxies (see references), with the
best estimate values for the QSO hosts from this paper. The lines sketch
in passive evolution models for 1.0 and 0.1 Gyr starbursts at z=20. 

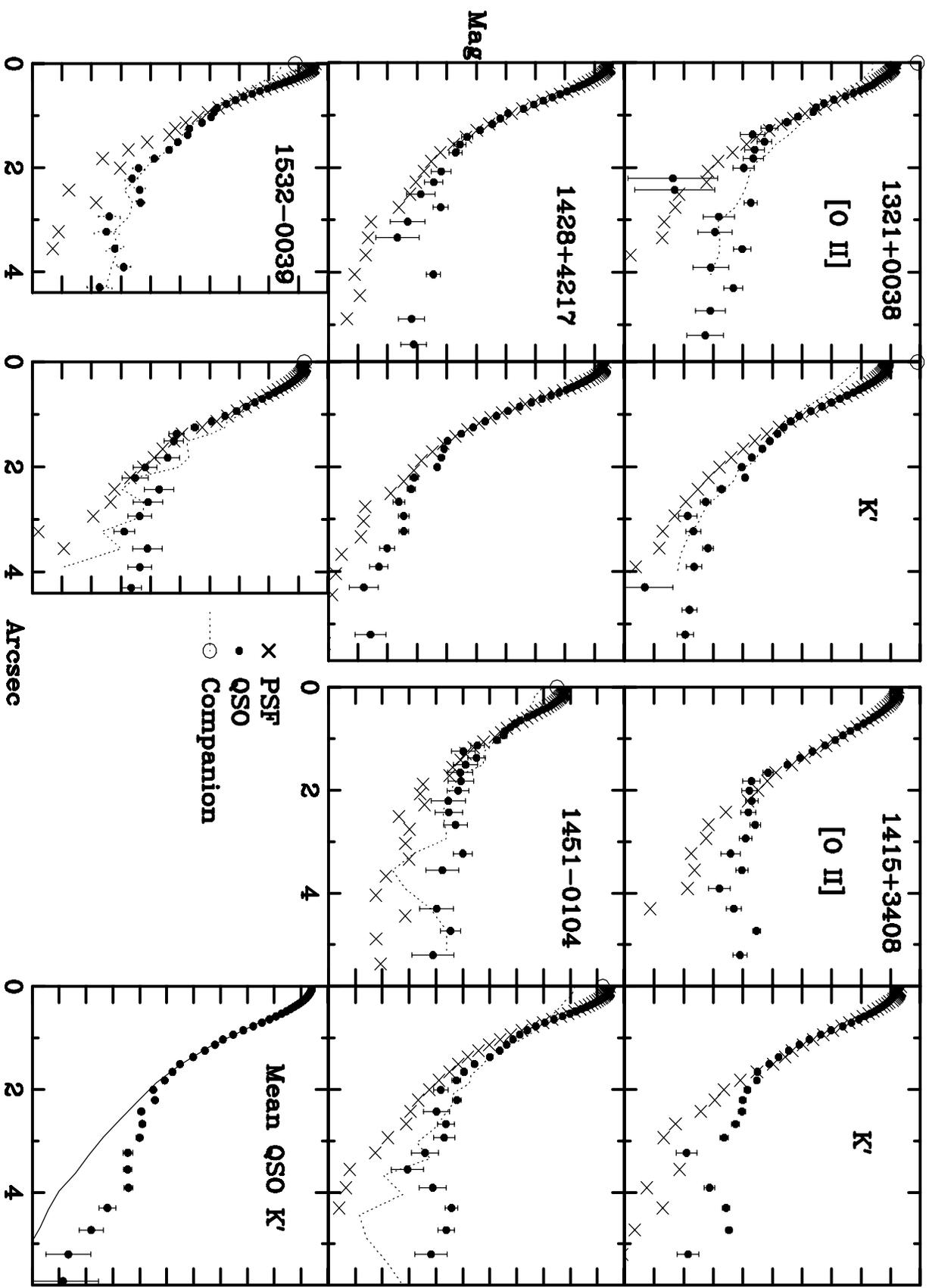
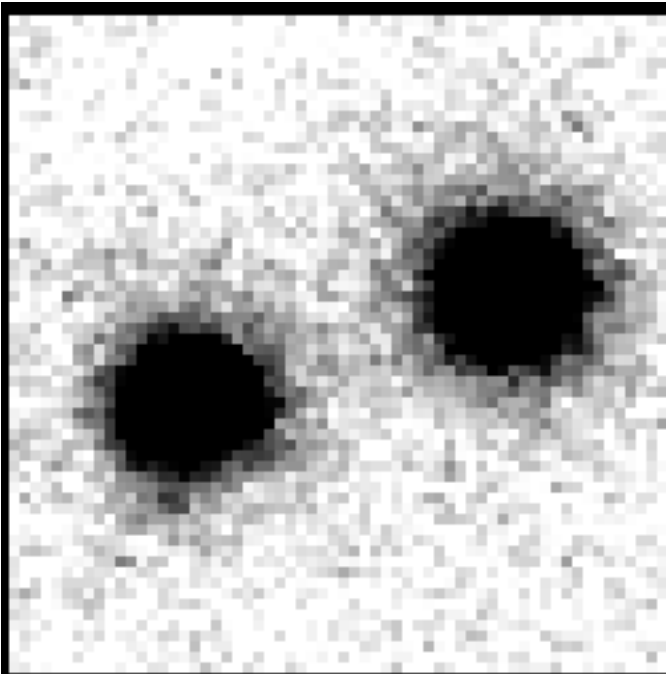
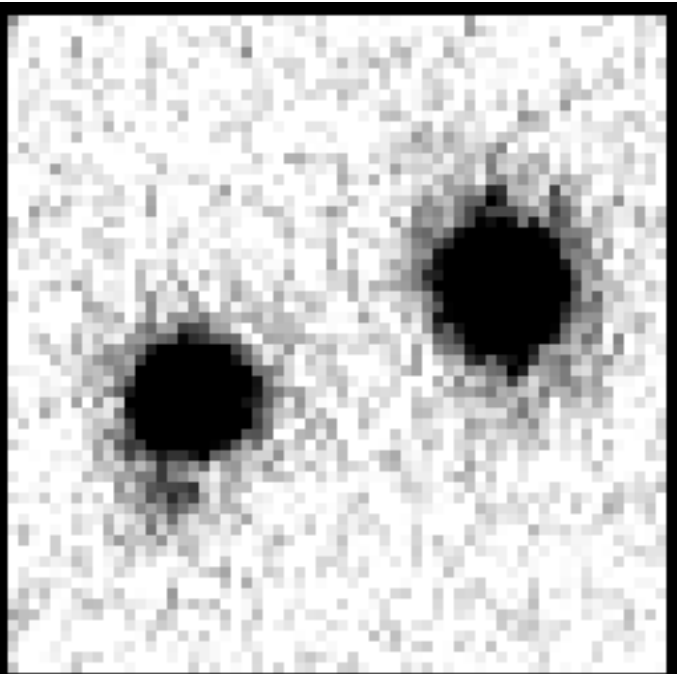
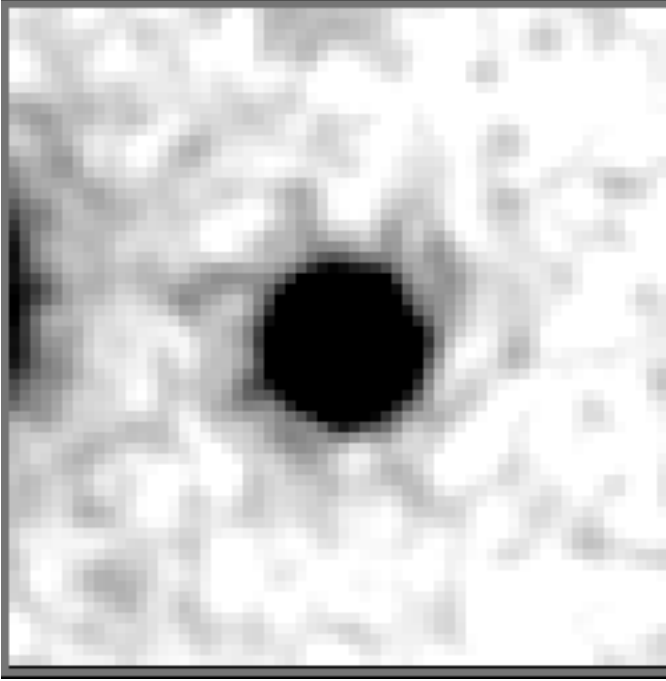
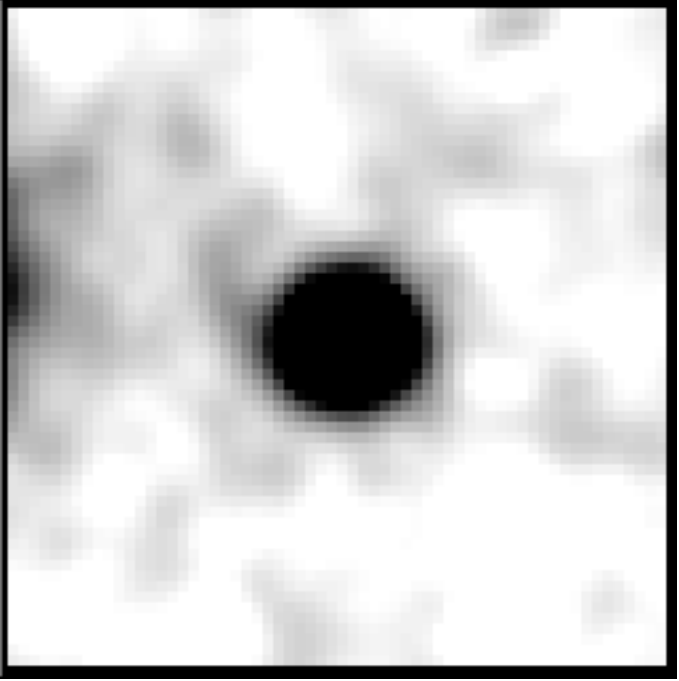
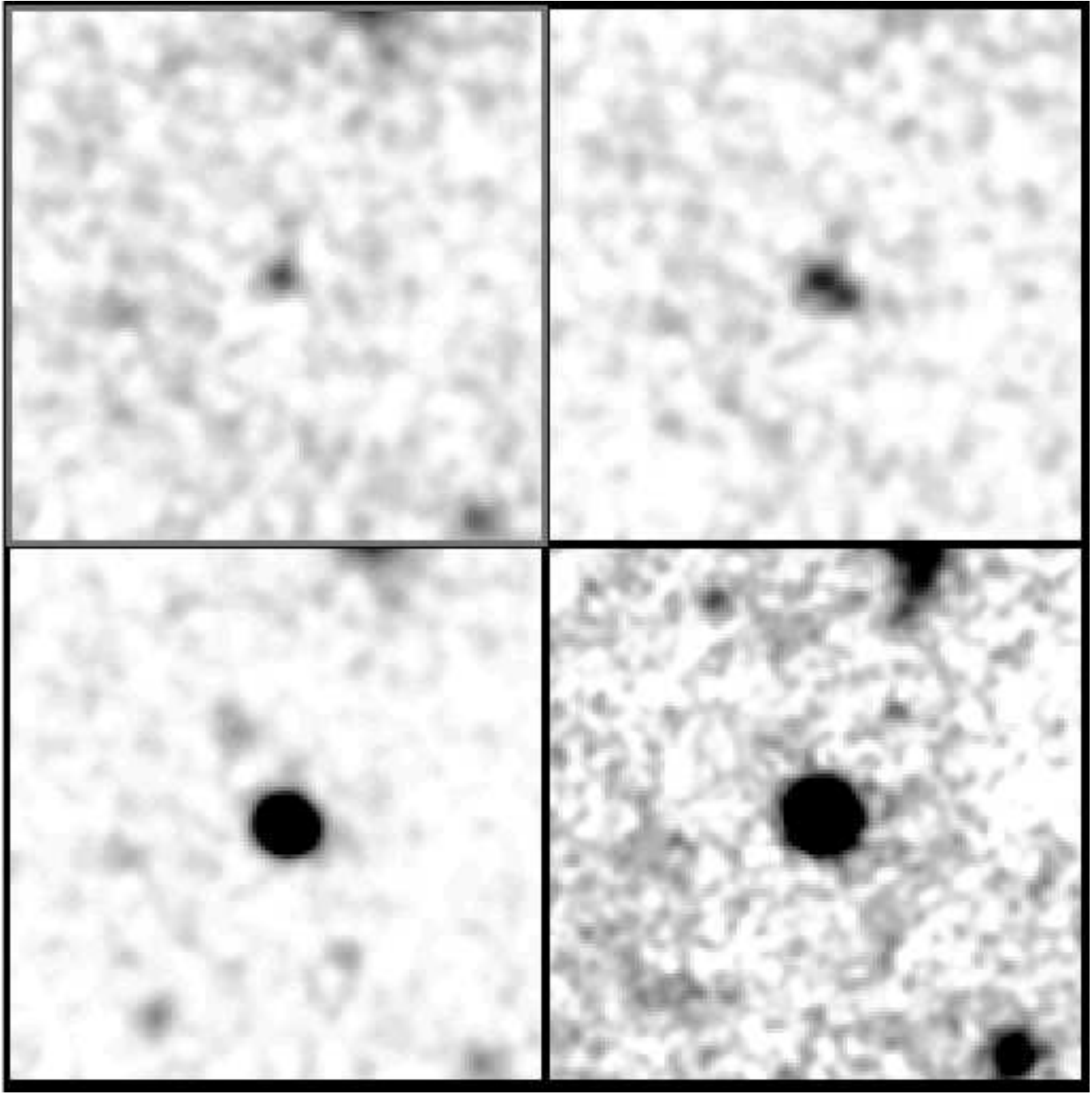
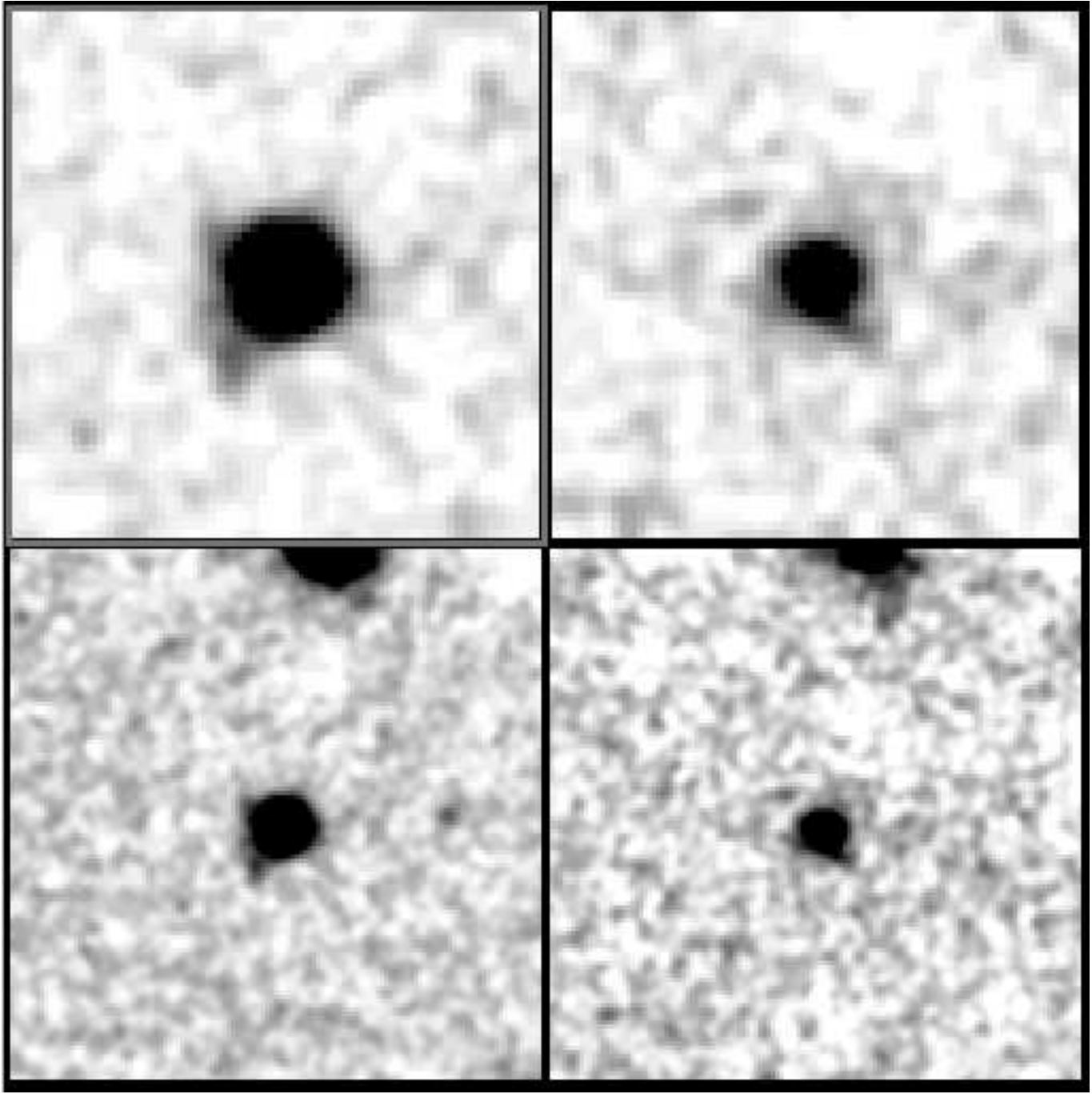
\end{document}